\documentstyle[aps,prl,graphicx,floats]{revtex}
\begin{document}

% SETS A DOUBLE SPACE
%%%%%\baselineskip=18pt

\title{{\bf Global persistence exponent of the two-dimensional
            Blume-Capel model}}
\author{Roberto da Silva{\bf
\thanks{E-mail: rsilva@dfm.ffclrp.usp.br}, }
        Nelson A. Alves{\bf
\thanks{E-mail: alves@quark.ffclrp.usp.br}, and }
        J. R. Drugowich de Fel\'{\i}cio{\bf
\thanks{E-mail: drugo@usp.br}}}
\address{{\it Departamento de F\'{\i}sica e Matem\'{a}tica,
     FFCLRP Universidade de S\~{a}o Paulo, Avenida Bandeirantes 3900,}\\
CEP 014040-901, \thinspace\ Ribeir\~{a}o Preto, S\~ao Paulo, Brazil}
\date{May, 14, 2002}
\maketitle

\begin{abstract}
The global persistence exponent $\theta_g$ is calculated for the 
two-dimensional Blume-Capel model following a quench to the critical 
point from both disordered states and such with small initial magnetizations.
 Estimates are obtained for the nonequilibrium critical dynamics
on the critical line and at the tricritical point.
 Ising-like universality is observed along the critical line
and a different value $\theta_g =1.080(4)$ is found 
at the tricritical point.

\end{abstract}



\vskip1.0cm

%%%%%%%%%%%%%%%%%%%%%%%%%%%%%%%%%%%%%%%%%%%%%%%%%%%%%%%%%%%%%%%%%%%%%%% 
%%%%%%%%%%%%%%%%%%%%%%%%%%%%%%%%%%%%%%%%%%%%%%%%%%%%%%%%%%%%%%%%%%%%%%%
%%\section{INTRODUCTION}
%%%%%%%%%%%%%%%%%%%%%%%%%%%%%%%%%%%%%%%%%%%%%%%%%%%%%%%%%%%%%%%%%%%%%%% 
%%%%%%%%%%%%%%%%%%%%%%%%%%%%%%%%%%%%%%%%%%%%%%%%%%%%%%%%%%%%%%%%%%%%%%% 
\vspace*{-0.5pt}
\indent
 Recent advances in nonequilibrium physics have enlarged our understanding
of critical phenomena. 
 As shown by Janssen {\it et al.} \cite{Janssen1} and Huse \cite{Huse},
universality and scaling behavior are already present in systems since 
the early stages of their relaxation processes after quenching from high 
temperatures to the critical one.
 This evolution is governed by a new independent exponent
$\theta$.
 Systems characterized by a nonconserved order parameter,
model A in the terminology of Halperin {\it et al.} \cite{HH1},
are described by a scaling function soon after a microscopic time scale
$t_{mic}$. Its general form for the $k{\rm th}$ moment of the magnetization
(e.g., in a ferromagnet) reads as
\begin{equation}
 M^{(k)}(t,\tau ,L,m_{0})=
 b^{-k\beta /\nu } M^{(k)}(b^{-z}t,b^{1/\nu }
 \tau,b^{-1}L,b^{x_{0}}m_{0}) \, .                        \label{magk}
\end{equation}
Here $b$ is an arbitrary spatial scaling factor, $t$ is the time evolution
and $\tau $ is the reduced temperature, $\tau =(T-T_{c})/T_{c}$. As usual,
the exponents $\beta $ and $\nu $ are the well-known static exponents, whereas
$z$ is the dynamic one. Equation (\ref{magk}) depends on the initial
magnetization $m_{0}$ and gives origin to the new exponent $x_{0}$,
scaling dimension of the initial magnetization, related to
$\theta$ by $x_{0}=\theta z +\beta /\nu$. 

 The nonequilibrium short-time exponent $\theta$ can be obtained
at the critical temperature from the scaling form of the first moment
\cite{Janssen1,Sch97,Review}, $ M(t,m_{0})\sim m_{0}t^{\theta }$. 
Usually, this exponent assumes a positive value corresponding 
to a critical initial slip, which is related to the anomalous behavior
of the magnetization when the system is quenched to $T_c$. 
  Numerical works indicate negative values for $\theta$ 
at the tricritical point \cite{SAF2002b} of the $S=1$ Blume-Capel 
model \cite{Blume66,BEG,Lawrie} in two dimensions ($2d$)
as well as for the Baxter-Wu \cite{Everaldo},
 Ising model with three-spin interactions
\cite{Claudia01} and for the four-state Potts model \cite{daSilva3}.
 The indication of a negative value for this exponent 
was theoretically deduced  by Janssen and Oerding \cite{Janssen2}
from a study of non-equilibrium relaxation at a
tricritical point.
 Numerical simulations of the $S=1$ Blume-Capel model \cite{SAF2002b} 
present for the dynamical
exponents the values $\theta=-0.53(2)$ and $z=2.21(2)$
at the tricritical point and values compatible
with heat-bath dynamics for the two-dimensional Ising model
($\theta=0.191(1)$ \cite{Sch97,Grass95}, 
$z=2.156(2)$ \cite{SAF2002a}) on the critical line.

 Under the same nonequilibrium conditions, a second critical exponent
has been presented in the literature \cite{Mglobal96}:
the global persistence exponent $\theta_g$. 
 It is related to the probability $P(t)$ that the global order
parameter has not changed sign up to time $t$ after a quench to $T_c$,
$P(t) \sim t^{-\theta_g}$.         
 This exponent has emerged from the concept of the local 
persistence phenomena in coarsening dynamics at $T=0$ 
\cite{Derrida94,Bray94,Stauffer94,Derrida95,Majumdar96,Mlocal96}.
%% The local persistence, defined as the fraction $P_l(t)$
%%of sites where after quenching from a high temperature to $T=0$ 
%%the local field has not changed sign up to time $t$
%%also decays as a power law characterized by a different exponent $\theta_l$,
%%$P_l(t) \sim t^{-\theta_l}$.         
 As argued by Majumdar {\it et al.} \cite{Mglobal96}, if
the dynamics of the global order parameter is described by a 
Markovian  process, $\theta_g$ is not a new independent exponent 
and it can be related to other critical exponents,
\begin{equation}
     \theta_g z=\lambda -d+1-\eta /2 \,.     \label{global}
\end{equation}
 Here $\lambda $ is the nonequilibrium exponent of the auto-correlation 
function \cite{Janssen1,Huse},
\begin{equation}
 A(t) = \frac{1}{L^d} \left\langle \sum\limits_{i}
        S_{i}(0) S_{i}(t) \right\rangle \, \sim \, t^{-\lambda/z } \, ,
\end{equation}
 which is related to the short-time exponent,
$\lambda =d - \theta z $.
 Therefore, Eq. (\ref{global}) can be rewritten as 
\begin{equation}
  \theta_g z= -\theta z +\frac{d}{2} -\frac{\beta}{\nu} \,.  \label{globalb}
\end{equation}
 However, the time evolution of the order parameter is, in general, 
a non-Markovian process and $\theta_g$  turns out to
be a new independent critical exponent describing the
stochastic dynamic process toward the equilibrium.

Contrary to the local persistence, the global persistence has been
less studied. Results have only been reported
for the $n \rightarrow \infty$ limit of the $O(n)$ model
($\theta_g =(d-2)/4$ for $2<d <4$ or $\theta_g=1/2$ for $d >4$) and 
to order $\epsilon= 4-d$ near $d=4$  
($\theta_g =1/2 -\epsilon (n+2)/(4n+32) + O(\epsilon^2)$), 
for $d=1$ Ising model ($\theta_g =1/4$) \cite{Mglobal96}, and 
$d=2$ Ising model \cite{Mglobal96,Stauffer96,SZ97}, 
as shown in Table I.
 As remarked in \cite{Mglobal96}, relation (\ref{global}) 
is satisfied for $n=\infty$ limit of the $O(n)$ model,
to first order in $\epsilon=4-d$, and also for $d=1$ Ising model. 
 However, for the $O(n)$ model, it has been shown \cite{Oerding97}
that the scaling relation (\ref{global}) for Markovian process is violated
at order $\epsilon^2$.

 In this paper we investigate the universality aspects of 
the global persistence exponent for the Blume-Capel model.
 This exponent is obtained by the straight application of
the power law behavior $P(t) \sim t^{-\theta_g}$         
and by means of time series data collapse.
 This study is also performed under different initial conditions:
random choices of all spins and sharp preparations of samples
with defined and nonzero magnetizations $m_0$ \cite{SZ97}.

The Blume-Capel \cite{Blume66} (BC) model is a spin-1 model
which has been used to describe the behavior of ${\rm ^3He}-{\rm ^4He}$
mixtures along the $\lambda$ line and near the critical mixing point.
Apart from its practical interest, the BC model has intrinsic interest
since it is the simplest generalization of the Ising model ($s=1/2$)
exhibiting a rich phase diagram with first and second-order transition
lines and a tricritical point.
 The Hamiltonian of the two-dimensional model is
\begin{equation}
 H=-J\sum\limits_{<i,j>}S_{i}S_{j}+D\sum\limits_{i=1}S_{i}^{2}\,,  \label{b1}
\end{equation}
where $<i,j>$ indicates nearest neighbors on $L^{2}$ lattices and
$S_{i}=\{-1,0,1\}$.
The parameter $J$ is the exchange coupling constant and $D$ is the
crystal field.
 We remark that along the critical line, this model presents a critical
behavior similar to the Ising model. However, exactly at the
tricritical point the exponents change abruptly. 
In \cite{Stilck} finite-size scaling combined with conformal invariance
permitted to observe a smooth change between Ising-like and
tricritical behavior.
 Ising-like behavior is reached only when $L\rightarrow \infty$, 
leading to the exact values of the Ising model critical exponents.

 In order to evaluate the persistence probability $P(t)$,
we define $\rho(t)$ as the fraction of samples which change their 
signals for the first time at the instant $t$.
 Our probability $P(t)$ is numerically calculated from the
cumulative distribution function such that the total magnetization does 
not cross the origin up to time $t$,
\begin{equation}
  P(t)=1-\sum\limits_{t'=1}^{t} \rho(t') \,.
\end{equation}
The spins $\{S_{i}\}$ are updated by the heat-bath algorithm and
our statistics rely on $N_B=5$ independent bins with $N_S=40000$ samples 
for $t$ up to 1000 MC sweeps and lattice size $L=80$.
 We quote estimates for the time intervals $[t_i, t_f]$ with the highest
values of goodness-of-fit $Q$ \cite{Press} 
for the linear regression obtained at every 10th measurement.
 
 Table II lists the points on the second order critical line and the 
tricritical point where we have performed our simulations.
 This table presents estimates for $\theta_g$ in function of different 
magnetizations $m_0$ to explore the effect of the initial configurations 
on the behavior of $P(t)$. 
 Linear extrapolations for $m_{0}\rightarrow 0$ are presented
in the last column.
 Here we follow the sharp preparation technique  to set 
the value $m_0$.
 Our typical time intervals for $m_0=0.0050$ correspond to 
$[100,500]$ (critical points) and  $[40,400]$ (tricritical point).
 Different time intervals with accepted values for $Q$,
present compatible results within our error estimates.

 Figures 1 and 2 illustrate, respectively, the decay of persistence 
probability for the specific critical point $D/J=0$, $k_{B}T/J=1.6950$ 
and for the tricritical point.

 Our simulations of the BC model
on the critical line reproduce (see Table II) the 
estimates obtained by Schulke and Zheng \cite{SZ97} for the 
$2d$ Ising model with  $m_0=0.0005$:
 $\theta_g=0.238(3)$ (with the HB dynamics) and 
 $\theta_{g}=0.236(3)$ (Metropolis algorithm).
  On the other hand, our simulations present the largest 
deviation (compared with the Ising model exponent)
for the largest initial magnetization $m_0=0.0050$, but
the expected universality is recovered as $m_0 \rightarrow 0$.
 It becomes clear the importance of the initial configurations
in measuring the persistence exponent.

 At the tricritical point we observe a faster decay of the
persistence probability ($\theta_g=1.080(4)$), 
characterizing a different persistence behavior.
 We repeat our simulations for two new critical points 
 \cite{Beale}
$D/J=1.87$, $k_{B}T/J= 0.800$ and $D/J=1.95$, $k_{B}T/J=0.650$,
closer to the tricritical one to follow this cross over effect. 
In this case, the persistence probability can be fitted by
a power law function only for short time intervals. 
Hence, 
% Our results identify a power law behavior for the persistence
%probability only for short time intervals at those new points. 
we have to restrict our analysis to shorter time intervals 
in order to obtain acceptable values for the goodness-of-fit. 
 We obtain for the first point $\theta_g=0.172(3)$  
in time interval $[80,400]$, while for the second point, closer 
to the tricritical one, we have to restrict further the interval to
$[100,300]$, leading to the estimate $\theta_g=0.345(5)$.
 Hence, this very different behavior characterized by those
new values of $\theta$ indicates 
a `transition' from the Ising-like 
$\theta_g \sim 0.23$ to the tricritical  $\theta_g \sim 1.08$
for finite lattices.

 The initial magnetization dependence of $P(t)$
can be cast in the following finite-size scaling relation \cite{Mglobal96},
\begin{equation}
P(t)=t^{-\theta_g}\,f(t/L^z) = L^{-\theta_g z}\,{\tilde f}(t/L^z)
                                                    \, ,  \label{majundar}
\end{equation}
which renders a different method to obtain the exponent $\theta_g$ from 
lattice sizes $L_1$ and $L_2$ \cite{Mglobal96}.
  For this end we define $W(t,L)=L^{\theta_g z}P(t)$, which turns out to 
be a function of $t/L^z$. 
  Therefore, if we fix the dynamic exponent $z$,
the exponent $\theta_g$ can be obtained by collapsing the time series
$W(t_2,L_2) = f(t_2/L_2^z)$ onto $W(t_1,L_1) = f(t_1/L_1^z)$ as follows.
  Under re-scaling, with $b=L_2/L_1$, $(L_2 > L_1)$, we obtain
\begin{equation}
    W(t_2,L_2) = \widetilde{W}(b^z t_1,b L_1) \, , 
\end{equation}
and the best estimate for $\theta_g$ corresponds to the minimization of 
\begin{equation}
\chi^{2}(\theta_g)=\sum\limits_t  
   \left( \frac{W(t,L)-\widetilde{W}(b^{z}t,bL)}
  {\left\vert W(t,L)\right\vert +
   \left\vert \widetilde{W}(b^{z}t,bL)\right\vert}\right)^{2} 
\end{equation}
by interpolating $\widetilde{W}$ to the time values $b^{z}t$.

This method is applied to the critical point $D/J=0,\, k_B T/J=1.6950$ 
and to the tricritical one.   
 We performed simulations with
lattice sizes $L=10,20,40$ and $80$ and initial disordered 
states to study the finite size dependence of $\theta_g$.
 Our simulations also rely on $N_{s}=40000$ samples and $N_{b}=5$ bins. 
 The collapse obtained from our largest pairs of lattices 
$(L_{1},L_{2})= (40,80)$ is displayed in Fig. 3 for the tricritical point.
A similar figure (not shown) is also obtained for the critical one.
 Results for the persistence exponent
are presented in Table III with the input values
 $z=2.106$ and $z=2.215$, 
respectively for the critical and  tricritical point \cite{SAF2002b}. 
 Here, we can observe a good agreement between both methods.
 However, we remark that the collapse method gives less precise results.

 Now, we return to Eq. (\ref{globalb}) to analyze the
markovian aspects of the time evolution of the magnetization.
 We verify that our numerical precision for
both the exponents \cite{SAF2002b}  $\theta$ 
 and $\theta_g$, evaluated on the critical line and at the tricritical point,
allows us to detect the non-markovian behavior. 
 Thus, these results substantiate the independence of 
global persistence and short-time exponents, characteristic of
a non-markovian dynamic evolution of the magnetization.

In summary, we have studied the effects of the initial magnetization 
on $P(t)$ and evaluated the global persistence exponent of the Blume-Capel 
model from different methods (linear extrapolation $m_{0}\rightarrow 0$
and collapse).
 The universality and independence 
of this dynamic exponent is explicitly shown for the BC model
along the critical line.
In addition, 
 the power law time dependence of the persistence probability
 is also exhibitted at the tricritical point, 
presenting a faster decay when compared with 
the Ising model exponent.

%%%%%%%%%%%%%%%%%%%%%%%%%%%%%%%%%%%%%%%%%%%%%%%%%%%%%%%%%%%%%%%%%%%%%%%%

\vspace{1cm}

%%%{\bf Acknowledgments} \vspace{0.4cm}

R. da Silva and J. R. Drugowich de Fel\'{\i}cio gratefully acknowledge
support by FAPESP and N.A. Alves by CNPq (Brazil).

%%%%%%%%%%%%%%%%%%%%%%%%%%%%%%%%%%%%%%%%%%%%%%%%%%%%%%%%%%%%%%%%%%%%%%%%%

\newpage 

%%%%%%%%%%%%%%%%%%%%%%%%%%%%%%%%%%%%%%%%%%%%%%%%%%%%%%%%%%%%%%%
% TABLE 1
\begin{table}[ht]
\renewcommand{\tablename}{TABLE}
\caption{\baselineskip=0.8cm 
         The global persistence exponent $\theta_g$ for the $2d$ Ising model
         with random and with pre-fixed small initial magnetization $m_0$.}
\begin{center}
\begin{tabular}{cccc}\\
\\[-0.3cm]
 Reference         & {\rm random} &$m_{0}=0.0010$ &$m_{0}=0.0005$ \\
\\[-0.35cm]
\hline
\\[-0.3cm]
~\cite{SZ97}       & 0.233(5)     & 0.237(5)     & 0.238(3) \\
~\cite{Stauffer96} &~0.225(10)    &              &   \\ 
~\cite{Mglobal96}  & 0.233(9)     &              &   \\
\end{tabular}
\end{center}
\end{table}
%%%%%%%%%%%%%%%%%%%%%%%%%%%%%%%%%%%%%%%%%%%%%%%%%%%%%%%%%%%%%%%

% TABLE 2
\begin{table}[ht]
\renewcommand{\tablename}{TABLE}
\caption{\baselineskip=0.8cm 
         The global persistence exponent $\theta_g$ from the 
         power law behavior for the $2d$ Blume-Capel model for
         different initial magnetizations $m_0$. Last column contains 
         our linear extrapolation to $m_0=0$.}
\begin{center}
\begin{tabular}{cccccc}\\
\\[-0.3cm]
D/J & $k_{B}T/J$  &$m_{0}=0.0050$ &$m_{0}=0.0025$ &$m_{0}=0.0005$ & 
                                                   extrapolated value \\
\\[-0.35cm]
\hline
\\[-0.3cm]
0      & 1.6950   & 0.219(3)      & 0.233(2)      & 0.237(3)    &0.241(4) \\
-3     & 2.0855   & 0.223(2)      & 0.233(1)      & 0.234(2)    &0.238(4) \\
-5     & 2.1855   & 0.223(2)      & 0.232(1)      & 0.235(1)    &0.237(2) \\
tricritical point &               &  \\
1.9655 & 0.610    & 1.054(3)      & 1.072(4)      & 1.076(3)    &1.080(4) \\
\end{tabular}
\end{center}
\end{table}
%%%%%%%%%%%%%%%%%%%%%%%%%%%%%%%%%%%%%%%%%%%%%%%%%%%%%%%%%%%%%%%

% TABLE 3
\begin{table}[ht]
\renewcommand{\tablename}{TABLE}
\caption{\baselineskip=0.8cm 
         The global persistence exponent $\theta_g$ for the best data 
         collapse for the $2d$ Blume-Capel model with
         initial magnetization $m_0=0$.}
\begin{center}
\begin{tabular}{cccc} \\
\\[-0.3cm]
    &    &{\rm ~critical point}   & {\rm ~tricritical point}  \\ 
    &    &{$(D/J=0,~k_B T/J=1.6950)$}  & {$(D/J=1.9655,~k_B T/J=0.610)$} \\ 
$L_1$ &$L_2$ & ~~~$\theta_g$      & ~~~$\theta_g$   \\    
\\[-0.35cm]
\hline
\\[-0.3cm]
10  & 20   &~~0.254(7)            &~ 0.88(2) \\
20  & 40   &~ 0.26(2)             &~ 1.08(2) \\
40  & 80   &~ 0.24(2)             &~ 1.06(2) \\
\end{tabular}
\end{center}
\end{table}
%%%%%%%%%%%%%%%%%%%%%%%%%%%%%%%%%%%%%%%%%%%%%%%%%%%%%%%%%%%%%%%%%%%%%%

\newpage

FIGURE CAPTIONS \\

Figure 1: Persistence probability $P(t)$ for the $2d$ Blume-Capel model
          for $L=80$ at a critical point with the sharp preparation of 
          the initial magnetization $m_0$. \\

Figure 2: Persistence probability $P(t)$ for the $2d$ Blume-Capel model
          for $L=80$ at the tricritical point with the sharp preparation of 
          the initial magnetization $m_0$. \\
 
Figure 3: Collapse of persistence probability at the tricritical point
          $D/J=1.9655,\, k_B T/J= 0.610$. \\

%%%%%%%%%%%%%%%%%%%%%%%%%%%%%%%%%%%%%%%%%%%%%%%%%%%%%%%%%%%%%%%

%FIGURE 1
\begin{figure}[!ht]
\begin{center}
\begin{minipage}[t]{0.95\textwidth}
\centering
\includegraphics[angle=-90,width=0.72\textwidth]{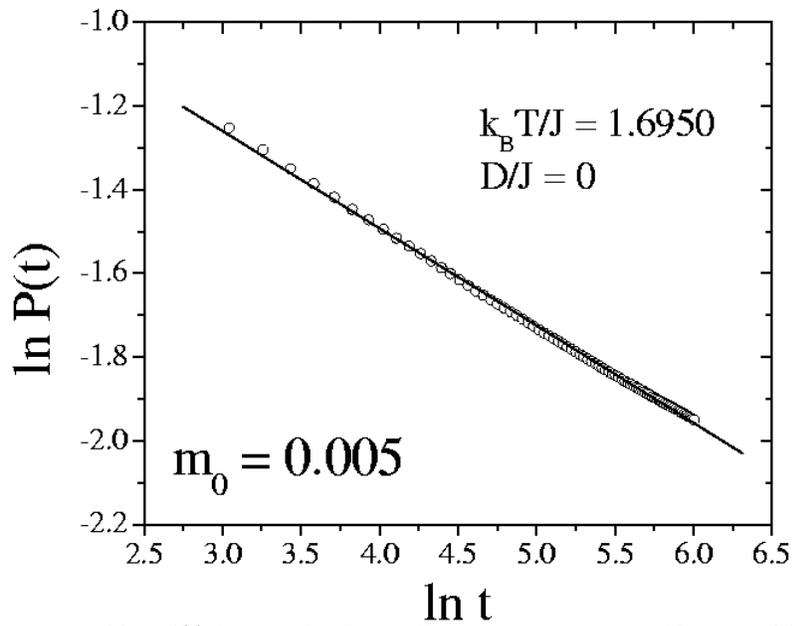}     
\caption{Persistence probability $P(t)$ for the $2d$ Blume-Capel model
          for $L=80$ at a critical point with the sharp preparation of 
          the initial magnetization $m_0$.} 
\label{Fig. 1}
\end{minipage}
\end{center}
\end{figure}

\newpage

%FIGURE 2
\begin{figure}[!ht]
\begin{center}
\begin{minipage}[t]{0.95\textwidth}
\centering
\includegraphics[angle=-90,width=0.72\textwidth]{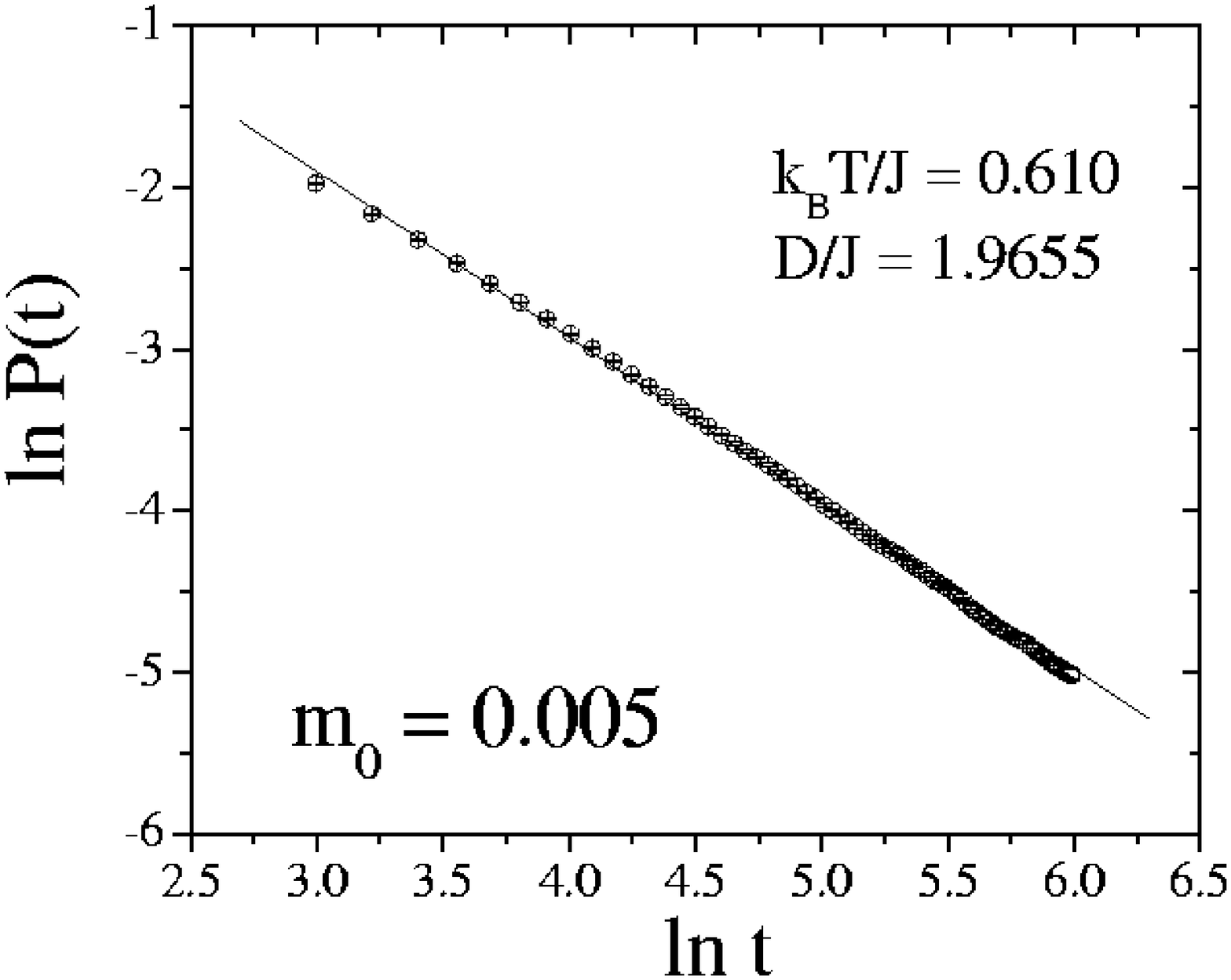}     
\caption{Persistence probability $P(t)$ for the $2d$ Blume-Capel model
          for $L=80$ at the tricritical point with the sharp preparation of 
          the initial magnetization $m_0$.}
\label{Fig. 2}
\end{minipage}
\end{center}
\end{figure}

%FIGURE 3
\begin{figure}[!ht]
\begin{center}
\begin{minipage}[t]{0.95\textwidth}
\centering
\includegraphics[angle=-90,width=0.72\textwidth]{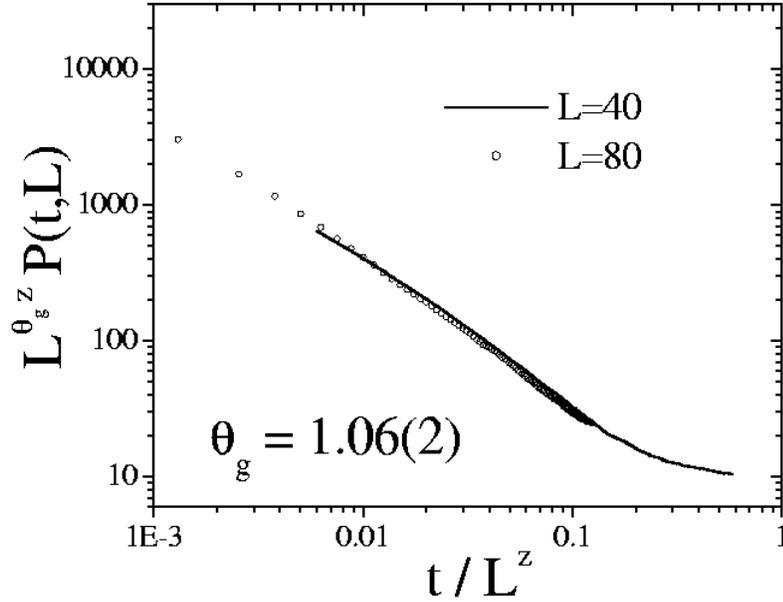}     
\caption{Collapse of persistence probability at the tricritical point
                   $D/J=1.9655,\, k_B T/J= 0.610$.}
\label{Fig. 3}
\end{minipage}
\end{center}
\end{figure}

\end{document}